# A Heuristic Method for Simplified Resource Allocation based on Comparative Advantage in Wireless Access Systems

Lin Cheng, *Senior Member*, *IEEE*
Next-Gen Systems
CableLabs
Louisville, Colorado, USA
l.cheng@cablelabs.com

Bernardo A. Huberman, *Fellow*, *APS*
Next-Gen Systems
CableLabs
Santa Clara, California, USA
b.huberman@cablelabs.com

*Abstract*—This paper presents a heuristic method for simplifying resource allocation in access systems, leveraging the concept of comparative advantage to reduce computational complexity while maintaining near-optimal performance. Using power-division non-orthogonal multiple access (PD-NOMA) as an example, we demonstrate how this approach mitigates the challenge of power allocation in multi-cell networks. Our method reduces the search space for optimization, significantly decreasing computational overhead while ensuring efficient spectrum utilization. In principle, the method reduces the dimensions of search space by half. Extensive analysis and simulations validate its effectiveness, highlighting its potential for practical deployment in next-generation wireless networks. The proposed framework can help streamline resource allocation in complex communication environments, enhancing system performance and scalability.

## I. INTRODUCTION

In wireless communications, advanced access technologies hold the promise of substantial improvements in overall network performance. These technologies are critical in addressing the rapidly growing demand for higher data rates, lower latency, and increased reliability, AI native interface, and co-existing sensing capability [1]. However, despite their theoretical advantages, the practical implementation of these access schemes remains a formidable challenge due to multiple technical and economic constraints [2].

One of the primary obstacles facing advanced access technologies is the sheer computational complexity involved in resource allocation and interference management. Unlike traditional orthogonal multiple access (OMA) schemes, which allocate dedicated time, frequency, or spatial resources to users, modern access methods such as non-orthogonal multiple access (NOMA), rate-splitting multiple access (RSMA), and integrated sensing and communication (ISAC) require intricate optimization techniques to efficiently allocate power and bandwidth while minimizing interference [3]-[5]. This complexity is exacerbated by the necessity for real-time adaptation in highly dynamic environments, where factors such as user mobility, fluctuating channel conditions, and varying network loads necessitate continuous recalibration of resource allocation policies [6].

Another significant challenge stems from hardware and energy efficiency constraints. Advanced access technologies often rely on sophisticated signal processing techniques, such as successive interference cancellation (SIC) in NOMA and full duplex or joint transmission and reception in coordinated multipoint (CoMP) schemes [7]. While these technologies theoretically improve network capacity, they impose substantial processing burdens on both base stations (BSs) and user equipment as well as backhaul, leading to increased power consumption and latency [8]. Furthermore, the implementation of these techniques at scale requires highly efficient hardware capable of handling the requirement on signal fidelity and computational overhead without introducing excessive digital processing, which remains a critical limitation in real-world deployments [9].

In addition, resource allocation in all access technologies is complicated by multiple intrinsic factors, including shared airtime among different BS-user pairs, the logarithmic relationship between power and efficiency (as dictated by Shannon's theorem), hierarchical network functions, and the presence of inter-cell interference (ICI) [10]-[13]. Prior research has explored innovative resource allocation strategies that can optimize the performance of wireless systems under realistic constraints, such as convex optimization, reinforcement learning, digital twin, and game-theoretic approaches, to address these challenges [14]-[19]. While these methods are powerful, they often further require significant computational resources and real-time adaptability. Resource allocation remains difficult to implement in large-scale, dynamic networks with advanced access technologies.

This study introduces a novel approach by leveraging the economic concept of comparative advantage [20], which has traditionally been used in trade theory, to streamline the search for optimal resource allocation. Comparative advantage suggests that entities should allocate resources where they have the most relative efficiency, and this heuristic principle can be applied to reduce the dimensionality of complex optimization problems. Instead of solving high-dimensional optimization problems directly, the comparative advantage principle can be utilized to identify a lower-dimensional subset of solutions. The integration of the economic theory into wireless resource allocation is a non-analytical solution and sub-optimal but has the potential to significantly simplify computational burdens while maintaining near-optimal performance. This method is inspired by similar heuristic approaches in economics, artificial intelligence, and computer networks, where reducing search space has been proven to improve efficiency without substantial performance trade-offs [21][22].

Specifically, we explore the application of comparative advantage as a heuristic tool for simplifying the resource allocation process in the example of power-division (PD)-NOMA. PD-NOMA has been widely investigated due to its ability to enhance spectral efficiency by enabling multiple users to share the same frequency and time resources [8]. However, on top of the challenges aforementioned, PD-NOMA encounters the complexity of SIC, error propagation issues, clustering optimization difficulties, and non-convex power allocation problems [15][23]. These challenges become even more pronounced in multi-cell networks, where ICI further complicates user clustering and power distribution problems. Several studies have analyzed the impact of ICI in multi-cell PD-NOMA and have proposed various clustering and power allocation schemes [10]-[12]. Despite these efforts, finding optimal solutions for power allocation in multi-cell scenarios remains an open problem, particularly when considering the computational feasibility of real-time implementation.

Through extensive analysis and simulation, we demonstrate that the proposed heuristic approach mitigates the computational complexity by reducing the dimensionality of the search space while maintaining statistically minor performance degradation in PD-NOMA systems. By incorporating comparative advantage, the proposed method provides a promising pathway for practical implementation of PD-NOMA, as well as a general framework for addressing the inherent complexities of advanced access technologies.

Furthermore, this method is not confined to PD-NOMA alone. The heuristic can be extended to other multiple access schemes, including OMA, CoMP, and hybrid access techniques, where resource allocation plays a critical role in determining overall system performance.

In the following sections, we will provide a detailed mathematical formulation of our approach, present simulation results that validate its effectiveness, and discuss its broader implications for next-generation wireless networks. Our findings suggest that by leveraging insights from economic theory, we can develop practical, low-complexity solutions for modern wireless communication challenges, thereby paving the way for more efficient and scalable deployments of advanced access networks.

## II. COMPARATIVE ADVANTAGE IN TRANSMISSION

In a cluster where there are $N$ users and $M$ base stations (BS), over a certain frequency band, BS-$j$ allocates a fraction $f_{i,j}$ of its power to be transmitted to user-$i$. The feasible and practical allocation set is

$$F = \{(f_{i,j})_{N \times M} : 0 \leq f_{i,j} \leq 1, \quad \sum_i f_{i,j} = 1\} \quad (1)$$

Hence the total received SINR at user-$i$ is

$$x_i = \sum_{1 \leq j \leq M} |h_{i,j}|^2 p_j f_{i,j} \quad 1 \leq i \leq N \quad (2)$$

where $h_{i,j}$ is the channel response between user-$i$ and BS-$j$ and $p_j$ is the power density transmitted by BS-$j$ normalized by the noise power density at user receivers. Later in (8) we will find this power variable $p_j$ does not affect the comparative advantage calculation and hence one can decouple power loading from this proposed method.

Denoting $x = (x_1 \ldots x_N)$ and assuming $x_i > 0$ (or otherwise a zero $x$ would exclude that user from the cluster and the problem becomes trivial), we define a target function

$$g(x) = \max_i \frac{w_i}{\log(1+x_i)} \quad (3)$$

that is proportional to the longest time consumption among users to finish transmission when the transmission to different users is independent, where $w_i$ is a relative coefficient that regulates the overall throughput of user-$i$, e.g. service level agreement (SLA).

In general, we want to minimize $g$ with respect to $x$, i.e.

$$\operatorname*{argmin}_x \max_i \frac{w_i}{\log(1+x_i)} \quad (4)$$

Although $g(x)$ is not continuous, around zero it (to be accurate, its opposite) satisfies the Inada conditions, especially

$$\lim_{x_i \to 0} \frac{\partial g(x)}{\partial x_i} = -\infty \quad 1 \leq i \leq N \quad (5)$$

Note that $x_i > 0$, as previously defined.

Also, searching $x$ is equivalent to searching $f = (f_{i,j})_{N \times M} \in F$, considering $h_{i,j}$ and $p_j$ are measured and predefined.

$$\min_x g(x) = \min_{f \in F} g \begin{pmatrix} \sum_{1 \leq j \leq M} |h_{1,j}|^2 p_j f_{1,j}, \\ \sum_{1 \leq j \leq M} |h_{2,j}|^2 p_j f_{2,j}, \\ \ldots \\ \sum_{1 \leq j \leq M} |h_{N,j}|^2 p_j f_{N,j} \end{pmatrix} \quad (6)$$

The search space $F$ of this minimum has a dimension of $(N-1) \times M$.

Now let's check $|h_{i,j}|^2 p_j$, the coefficients of $f_{i,j}$ in $x_i$. For any BS-$j_1$ and $j_2$ and user-$i_1$ and $i_2$, assumably we observe the following criterion

$$\frac{|h_{i_1,j_1}|^2 p_{j_1}}{|h_{i_1,j_2}|^2 p_{j_2}} > \frac{|h_{i_2,j_1}|^2 p_{j_1}}{|h_{i_2,j_2}|^2 p_{j_2}} \quad (7)$$

or simply

$$\left|\frac{h_{i_1,j_1}}{h_{i_1,j_2}}\right| > \left|\frac{h_{i_2,j_1}}{h_{i_2,j_2}}\right| \tag{8}$$

According to comparative advantage, this criterion tells us one of $f_{i_1,j_2}$ and $f_{i_2,j_1}$ must be zero when the minimum of $g$ is achieved. This is proved in [24] with trivial modifications, e.g. we search for the minimum instead of the maximum of $g$ and the coefficients $|h_{i,j}|^2 p_j$ are equivalent to $b_{j,i}$ in [24].

In a very practical simplification, we look at two BSs with an inter-cell region and $N$ users (note that only two BSs can achieve full-rate block-coded CoMP transmission). Without loss of generality, we can order the users by comparative advantage so that BS-1 has comparative advantage in serving users with smaller indices, i.e. the criterion of

$$\left|\frac{h_{1,1}}{h_{1,2}}\right| > \cdots > \left|\frac{h_{N,1}}{h_{N,2}}\right| \tag{9}$$

By the comparative advantage characterization, for any $1 \leq i_1 < i_2 \leq N$ it must be that either $f_{i_1,2} = 0$ or $f_{i_2,1} = 0$. Therefore, there must exist some $I$ such that

$$\begin{aligned} f_{i,1} > 0, f_{i,2} = 0 & \quad \text{for } 1 \leq i < I \\ f_{i,1} = 0, f_{i,2} > 0 & \quad \text{for } I < i \leq N \end{aligned} \tag{10}$$

In words, BS-1 should transmit to user $1 \ldots I-1$ and possibly $I$, and BS-2 should transmit to user $I+1 \ldots N$ and possibly $I$.

This effectively reduces the dimension of the search space from $2N - 2$ down to $N - 1$.

However, one may have noticed that (3) assumes independent reception among users, while in reality the expression of $x$ with respect to $f$ is more complex considering SIC, ICI, and non-full-rate CoMP. In fact, in (3) each element in $g(x)$ before the $\max(\ )$ operation is differentiable with respect to $f_{i,j}$ and a closed form of solution is not impossible in our "over-simplified" assumption of independent reception.

We make this assumption because an accurate explicit form of the criterion for comparative advantage characterization is impractical. In the next section, we will analyze how the assumption over-simplifies the problem and then we will further see that using the criterion from an over-simplified model, although may not capture all the nuances of the actual communication system, can still keep the estimation of comparative advantage statistically feasible, guide the optimization process effectively, and hence only cause rare outliers in optimization results.

## III. NOMA AS AN EXAMPLE

In non-independent multi-cell multi-user transmission, with NOMA and full-rate CoMP included, the limiting SINR, i.e. the SINR that is used to calculate bit rate, at user-$i$ can be expressed as

$$\eta_{\tau,i} = \min \left\{ \begin{array}{c} \frac{x_i}{\sum_{k \in \mathbf{A}_{\tau,i}} \sum_{1 \leq j \leq M} |h_{i,j}|^2 p_j f_{k,j} + 1}, \\ \min_{k \in \mathbf{A}_{\tau,i}} \frac{\sum_{1 \leq j \leq M} |h_{k,j}|^2 p_j f_{i,j}}{\sum_{l \in \mathbf{A}_{\tau,i}} \sum_{1 \leq j \leq M} |h_{k,j}|^2 p_j f_{l,j} + 1} \end{array} \right\} \tag{11}$$

where $\mathbf{A}_{\tau,i}$ is the set of users that come after user-$i$ in SIC decoding order $\tau$, i.e. all users in $\mathbf{A}_{\tau,i}$ need to decode user-$i$'s signal for SIC. The first term inside min{ } is the SINR at user-$i$, while the second term is the lowest SINR of user-$i$'s signal across all users that come after user-$i$ in SIC decoding order $\tau$. This expression includes all downlink scenarios [10][11]. Note that the min( ) operations have to be included even if the SIC decoding order is determined, due to the nature of multi-cell [11][12].

Replacing $x$ by $\eta_\tau = (\eta_{\tau,1} \ldots \eta_{\tau,N})$ in (3), the target function becomes

$$g(\eta_\tau) = \max_i \frac{w_i}{\log(1+\eta_{\tau,i})} \tag{12}$$

It still satisfies the Inada conditions with respect to $\eta_\tau$. However, since $\eta_\tau$ is not linear to $f$ and also subject to the SIC order which changes its expression with respect to $f$, we cannot explicitly find the coefficients to express the criterion for comparative advantage characterization, let alone giving a closed form of the solution ($\frac{\partial g}{\partial f_{i,j}}$ is not differentiable). Instead, we keep using the criterion (10) as a replacement to approximate comparative advantage.

Let's use two BSs and two users as an example. From (11) we have

$$\begin{aligned} \eta_{\tau_0,1} &= \frac{|h_{11}|^2 p_1 f_{11} + |h_{12}|^2 p_2 f_{12}}{|h_{11}|^2 p_1 f_{21} + |h_{12}|^2 p_2 f_{22} + 1} \\ \eta_{\tau_0,2} &= \frac{|h_{21}|^2 p_1 f_{21} + |h_{22}|^2 p_2 f_{22}}{|h_{21}|^2 p_1 f_{11} + |h_{22}|^2 p_2 f_{12} + 1} \\ \eta_{\tau_1,1} &= \min \left\{ \begin{array}{c} \frac{|h_{11}|^2 p_1 f_{11} + |h_{12}|^2 p_2 f_{12}}{|h_{11}|^2 p_1 f_{21} + |h_{12}|^2 p_2 f_{22} + 1}, \\ \frac{|h_{21}|^2 p_1 f_{11} + |h_{22}|^2 p_2 f_{12}}{|h_{21}|^2 p_1 f_{21} + |h_{22}|^2 p_2 f_{22} + 1} \end{array} \right\} \\ \eta_{\tau_1,2} &= |h_{21}|^2 p_1 f_{21} + |h_{22}|^2 p_2 f_{22} \\ \eta_{\tau_2,1} &= |h_{11}|^2 p_1 f_{11} + |h_{12}|^2 p_2 f_{12} \\ \eta_{\tau_2,2} &= \min \left\{ \begin{array}{c} \frac{|h_{21}|^2 p_1 f_{21} + |h_{22}|^2 p_2 f_{22}}{|h_{21}|^2 p_1 f_{11} + |h_{22}|^2 p_2 f_{12} + 1}, \\ \frac{|h_{11}|^2 p_1 f_{21} + |h_{12}|^2 p_2 f_{22}}{|h_{11}|^2 p_1 f_{11} + |h_{12}|^2 p_2 f_{12} + 1} \end{array} \right\} \end{aligned} \tag{13}$$

We incorporate SIC orders into the equation and accordingly we optimize the minimum of three target functions corresponding to different SIC orders, i.e.

$$\min\{g(\eta_{\tau_0}), g(\eta_{\tau_1}), g(\eta_{\tau_2})\} \tag{14}$$

where $\tau_1$ is SIC order that user-1 has a smaller index, $\tau_2$ is SIC order that user-2 has a smaller index, $\tau_0$ is without SIC.

As aforementioned, we keep the criterion simple, and (9) and (10) in this case can be rewritten as

$$\begin{cases} f_{12}f_{21} = 0 & \text{if } \left|\frac{h_{11}}{h_{21}}\right| > \left|\frac{h_{12}}{h_{22}}\right| \\ f_{11}f_{22} = 0 & \text{else} \end{cases} \quad (15)$$

This describes a subspace of $F$ with only one instead of two dimensions. Then we search in this subspace for the minimum of the target functions.

In addition, we define a variable that quantifies normalized comparative advantage

$$\alpha = \left|\frac{|h_{1,1}h_{2,2}|^2 - |h_{1,2}h_{2,1}|^2}{|h_{1,1}h_{2,2}|^2 + |h_{1,2}h_{2,1}|^2}\right| \quad (16)$$

Using these expressions, below we will run Monte Carlo simulations and see the penalty induced by reducing the search space of $F$ based on comparative advantage is statistically minimal. Moreover, we will also see that $\alpha$ works as an indicator of a global optimum lying in the reduced search space.

IV. NUMERICAL RESULTS

To examine the feasibility of the proposed method, we run a Monte Carlo simulation where we uniformly and randomly cast two users over a 1000-meter-by-500-meter area (horizontal: 0-1000 m; vertical: 0-500 m). Two BSs are located at coordinates of (250, 250) and (750, 250) [11]. The two users can communicate with either or both of the BSs at a center frequency of 1 GHz. The shared channel has a bandwidth of 20 MHz. The transmitted power from each BS is at 10 W [11]. The wireless channels are slow fading channels, i.e. channel response does not change within one allocation cycle, with a path loss exponent of 3.

In Fig. 1, the optimal allocation of each instance corresponds to one point on the two-dimension region. The entire region other than the four corner points in Fig. 1 stand for simultaneous downlink transmission to two users and have to be realized by either OMA or NOMA. In this study, we focus on NOMA, as it on average outperforms OMA in our setup. These optimal points, as references, are derived from brutal force searches over the entire two-dimension region in Fig. 1. Note that the target function, as aforementioned, is not continuous over the region. As the proposed method describes, instead of searching over the entire region, we search the subspace limited by (15), i.e. two of the four edges in Fig. 1, selected by comparative advantage:

$$f_{1,1} = 0 \,\&\, 0 \le f_{1,2} \le 1 \cup 0 \le f_{1,1} \le 1 \,\&\, f_{1,2} = 1$$
$$if \left|\frac{h_{1,1}}{h_{1,2}}\right| < \left|\frac{h_{2,1}}{h_{2,2}}\right| \quad (17)$$

$$f_{1,1} = 1 \,\&\, 0 \le f_{1,2} \le 1 \cup 0 \le f_{1,1} \le 1 \,\&\, f_{1,2} = 0$$
$$if \left|\frac{h_{1,1}}{h_{1,2}}\right| > \left|\frac{h_{2,1}}{h_{2,2}}\right| \quad (18)$$

Results show that in 90% of the 10,000 instances the proposed method converges to the global optimum regardless of its limited search space, i.e. reducing the search space from two-dimension to one-dimension, when $w_1 = w_2$, and 92% when $w_1 = 2w_2$. We also change the noise power and see how it impacts this percentage, as shown in Fig. 2. In general, a noisier channel/receiver or a lower SNR leads to greater alignment between the suboptimum provided by the proposed method and the global optimum.

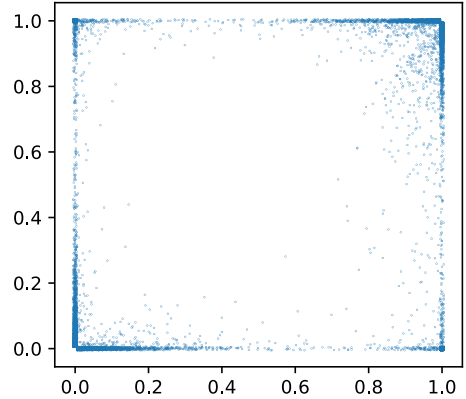

Fig. 1. Optimal power allocation of 10,000 random instances of a two-BS two-user system. x- and y-axis are $f_{1,1}$ and $f_{1,2}$ respectively. Noise power at receivers is set to $5 \times 10^{-11}$ W. $w_1 = w_2$.

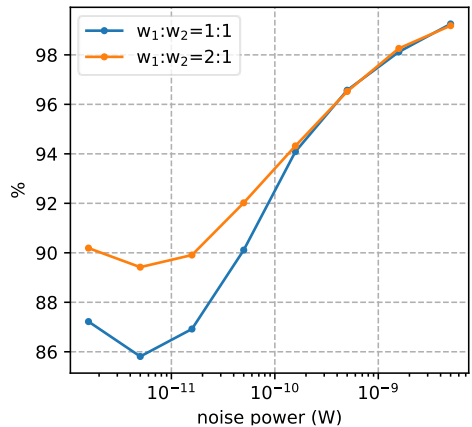

Fig. 2. Receiver noise power vs percentage of simulation instances where proposed method gives global optimum.

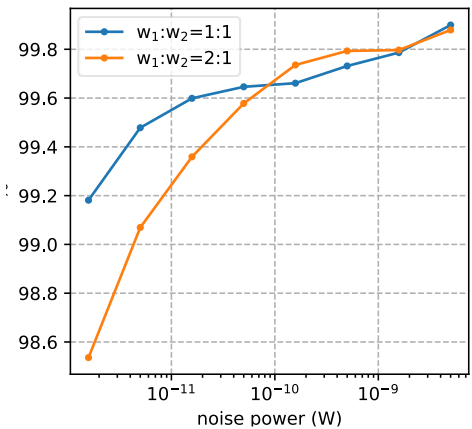

Fig. 3. Receiver noise power vs percentage of simulation instances that have global optimums on four edges and the results given by proposed method are global optimums.

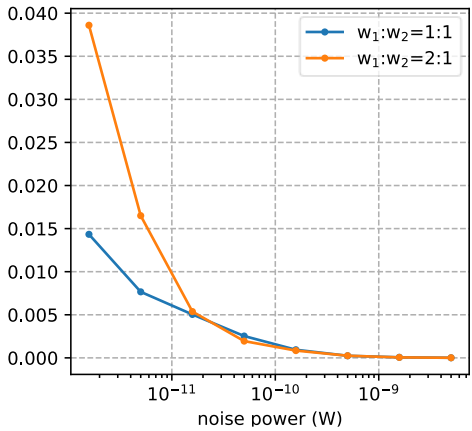

Fig. 4. Receiver noise power vs average degradation of the value of the target function at optimum provided by the proposed method compared with the value of the target function at global optimum.

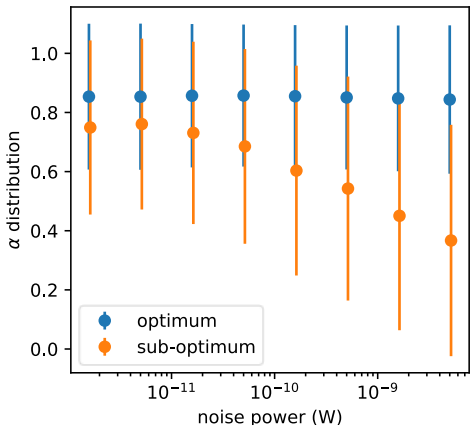

Fig. 5. Receiver noise power vs mean and standard deviation of $\alpha$ of the result given by proposed method being (blue) global optimum and (orange) sub-optimum. $w_1 = w_2$.

In fact, when a global optimum happens to lie on any of the four edges in Fig. 1, this global optimum can very likely be found by the proposed method (which only checks two of the four edges) and this likelihood is greater than 99.2% when $w_1 = w_2$, and greater than 98.5% when $w_1 = 2w_2$, over all the simulated, and practical, receiver noise powers, as shown in Fig. 3.

As we can see in Fig. 1, many optimal allocations are off from the edges hence cannot be included in the reduced sub-space described by the method. However, the values of the target function at these optimums are not far off from the values of the target function at the sub-optimums provided by the proposed method. In Fig. 4, we can see that the average degradation from the proposed method with respect to global optimum is less than 0.04% in worst scenario. This means that the method is an effective way to reduce the search space of power allocation with a minimal number of outliers.

In fact, in each instance we can use $\alpha$ in (16) to predict whether the optimization result given by the method is degraded from the global optimum. Fig. 5 shows the mean and standard deviation of $\alpha$ under two cases: (1) the result given by the method is exactly the global optimum, i.e. the global optimum lies inside the reduced search space; (2) the result given by the method is a sub-optimum, i.e. the global optimum lies outside the reduced search space. We can see that $\alpha$ can work as a differentiator between the two cases, especially at low SINR region. As a result, one can use $\alpha$ to decide whether to use the proposed method for each allocation instance. For example, when noise power is at $5\times10^{-9}$ W, the proposed method is more likely to give the global optimum when $\alpha$ is measured above 0.7, while more likely to give a sub-optimum with a low degradation when $\alpha$ is measured below 0.7.

## V. Discussion

### A. OMA

As aforementioned, PD-NOMA is solely an example to demonstrate our idea. In addition to power allocation in PD-NOMA, the proposed allocation method can also be applied to frequency/time resources in OMA systems. Simulation results (not included in this paper due to page limitation) indicate its effectiveness in these systems as well. The results also indicate that the proposed method, when implemented in OMA, has a nearly 100% probability to find the global optimum while reducing the dimensions of the search space by half.

### B. Target Function

It is important to clarify that the target functions presented in (3) and (12) serve merely as an illustrative example. Depending on the specific application scenario, one has the flexibility to select any appropriate target function based on the goal of optimization. As a general guideline, the target function should reflect variations in system transmission time consumption, rather than other metrics such as average user throughput. Additionally, when considering independent transmissions, as discussed in II, the chosen target function must adhere to the Inada conditions.

In the case of (3) and (12), we assume no handover within one allocation cycle, i.e. $f_{i,j}$ is static. However, this may cause part of the power of a BS under-utilized towards the end of the cycle. In practical scenarios, a BS dynamically redistribute its power among users, e.g. BS-$j$ initially allocates $f_{i_1,j}$ to user-$i_1$ then after $i_1$ finishes $j$ switches $f_{i_1,j}$ to $i_2$ and $f_{i_2,j}$ becomes $f_{i_1,j} + f_{i_2,j}$. Such dynamic allocation reduces system transmission time compared with the static approach outlined in (3) and (12). Given the complexity of deriving an explicit target function for a large number of users, we conducted a simulation in a two-cell, two-user system with dynamic power allocation. Correspondingly the target function of such system is modified from (12) into

$$g_\tau^* =$$

$$\begin{cases} \dfrac{w_1}{\log(1+\eta_{\tau,1})} + \dfrac{w_2 - w_1\frac{\log(1+\eta_{\tau,2})}{\log(1+\eta_{\tau,1})}}{\log\left(1+p_1|h_{2,1}|^2+p_2|h_{2,2}|^2\right)} & \text{if } \dfrac{w_1}{\log(1+\eta_{\tau,1})} < \dfrac{w_2}{\log(1+\eta_{\tau,1})} \\ \dfrac{w_2}{\log(1+\eta_{\tau,2})} + \dfrac{w_1 - w_2\frac{\log(1+\eta_{\tau,1})}{\log(1+\eta_{\tau,2})}}{\log\left(1+p_1|h_{1,1}|^2+p_2|h_{1,2}|^2\right)} & \text{else} \end{cases}$$

(19)

With other procedures kept the same, the simulation results show observations similar to the ones in IV.

*C. Optimization*

In III, the NOMA example explores power allocation between BSs and users, as well as SIC scenarios, for a single resource block (RB) or the entire bandwidth. For a more comprehensive optimization, an additional layer of BS-RB power allocation should be considered. This would involve dividing the bandwidth into multiple RBs, determining the total transmission power from each BS for each RB, and then iteratively applying the current NOMA algorithm to achieve a more global optimum. Such a search is expected to outperform both NOMA and OMA individually, as it encompasses both as special cases. However, this approach is significantly more complex. Nevertheless, the proposed method works as an independent step in reducing the dimensionality of the search space despite the increased complexity.

## VI. CONCLUSION

In conclusion, our proposed method of using comparative advantage as a heuristic offers an alternative perspective on the problem of searching optimal allocation. By significantly reducing the search space while maintaining performance, the method effectively addresses the complexities associated with decoding and ICI in the networks. For typical two-cell systems, we have demonstrated that the method reduces the dimension of the search space from $2N - 2$ to $N - 1$. The simulation results indicate that the method converges to near-optimal solutions with minimal degradation, showcasing its feasibility for practical low-computation implementation. Future research should explore the potential of generalizing the method with regard to modulation formats and the integration of our approach within broader optimization frameworks to further enhance system performance in evolving wireless environments.